\newif\ifconfpaper
\confpapertrue 

\newif\ifconfletter

\ifconfpaper
\documentclass[letterpaper, 10 pt, conference]{ieeeconf}
\IEEEoverridecommandlockouts
\fi

\ifconfletter
\documentclass[letterpaper, 10 pt, journal, twoside]{IEEEtran}  
\fi

\usepackage{graphicx}
\usepackage{amsmath}
\usepackage{amsfonts}
\usepackage{bm}
\usepackage{booktabs}

\newcommand{\secref}{Section~\ref}
\newcommand{\figref}{Figure~\ref}
\newcommand{\tabref}{Table~\ref}

\ifconfletter
\begin{document}
\fi
%
\ifconfletter
\title{Agreeing to Interact in Human-Robot Interaction using Large Language Models and Vision Language Models}
\else
\title{\LARGE \bf Agreeing to Interact in Human-Robot Interaction using Large Language Models and Vision Language Models}
\fi
%
%
%

\ifconfletter
\author{Kazuhiro Sasabuchi$^{1}$, Naoki Wake$^{1}$, Atsushi Kanehira$^{1}$, Jun Takamatsu$^{1}$, and Katsushi Ikeuchi$^{1}$%
\thanks{Manuscript received: Month, dd, year; Revised Month, dd, year; Accepted Month, dd, year.}
\thanks{This paper was recommended for publication by Editor XXX upon evaluation of the Associate Editor and Reviewers' comments.
}
\thanks{$^{1}$All authors are with Applied Robotics Research, Microsoft, Redmond, WA, USA
        {\tt\footnotesize Kazuhiro.Sasabuchi@microsoft.com}}%
\thanks{Digital Object Identifier (DOI): see top of this page.}
}
\else
\author{Kazuhiro Sasabuchi$^{1}$, Naoki Wake$^{1}$, Atsushi Kanehira$^{1}$, Jun Takamatsu$^{1}$, and Katsushi Ikeuchi$^{1}$%
\thanks{$^{1}$All authors are with Applied Robotics Research, Microsoft, Redmond, WA, USA
        {\tt\small Kazuhiro.Sasabuchi@microsoft.com}}%
}
\fi
%
%

\ifconfletter
\markboth{IEEE Robotics and Automation Letters. Preprint Version. Accepted Month, year}
{Sasabuchi \MakeLowercase{\textit{et al.}}: The Title} 
\fi

%



\ifconfpaper
\begin{document}
\fi

\maketitle

\ifconfpaper
\thispagestyle{empty}
\pagestyle{empty}
\fi

\begin{abstract}
In human-robot interaction (HRI), the beginning of an interaction is often complex. Whether the robot should communicate with the human is dependent on several situational factors (e.g., the current human's activity, urgency of the interaction, etc.). We test whether large language models (LLM) and vision language models (VLM) can provide solutions to this problem. We compare four different system-design patterns using LLMs and VLMs, and test on a test set containing 84 human-robot situations. The test set mixes several publicly available datasets and also includes situations where the appropriate action to take is open-ended. Our results using the GPT-4o and Phi-3 Vision model indicate that LLMs and VLMs are capable of handling interaction beginnings when the desired actions are clear, however, challenge remains in the open-ended situations where the model must balance between the human and robot situation.
\end{abstract}

\ifconfletter
\fi

%
\ifconfletter
\IEEEpeerreviewmaketitle
\fi

\section{Introduction}
\label{introduction}
%
%
%
%
\ifconfletter
\IEEEPARstart{T}{he} first sentence of the introduction.
\else
While many research on HRI assume that the human and the robot is in a state ready to interact with each other, beginning the human-robot interaction is often complex and cannot be ignored.
\fi
For example, when the human is waving at the robot, the robot can start interacting as there is a sign of acceptance. If the human is not noticing the robot, the robot should first try to catch the human's attention. However, if the human not noticing the robot is talking on a phone, the robot should instead wait for the phone call to end before speaking. The situation is even more complex if the robot had to report something urgent such as \textit{``there is a fire in the house,"} where the robot should after all try to catch the human's attention as the report is urgent.
Humans can understand these situational differences and take the appropriate action (speak, wait, or catch attention) but there remains a question on how to achieve such 
action decisions on a robot. 

In our past approach \cite{sasabuchi2018agreeing},
we have used human gaze patterns and a Hidden Markov Model to detect signs of human acceptance versus not noticing.
A heuristic mapping was then used to map the signs into robot actions such as catch attention or begin speaking.
Although the na\"ive approach has shown successful results, the model had limited ability to capture the situational context such as phone calls.

To this end, recent VLM are capable of providing a more high-level description of an image obtained from a robot's camera. 
In addition, recent LLM are capable of reasoning and provide answers to text-based questions \cite{binz2023using}. 
Thus, LLMs might provide reasonable answers to questions such as
\textit{``The robot is trying to report an urgent message but the person is talking over the phone. How should the robot act in this situation?"} with the expectation of the answer being \textit{``Although the person is talking over the phone, the robot should catch the person's attention by saying 'excuse me.' as the report is urgent."}

As a replacement to using a na\"ive model and heuristic mapping of actions, VLMs could help understand the situation; 
LLMs could further 
help generate the action the robot should take. Yet there remain several questions such as whether the above expectations hold true, how should the prompt be designed (what information about the interaction situation should be included) 
are there other ways of leveraging VLMs and LLMs for the problem (can the problem be solved directly from image input using a single VLM instead)? In this paper, we discuss these questions to understand how VLMs and LLMs can play in the problem of beginning a human-robot interaction.

The contributions of this paper are listed as follows:
\begin{itemize}
\item Introducing prompts and system design choices using VLM / LLM models to generate appropriate actions at the beginning of a human-robot interaction. 
\item A test set 
containing 84 situations. The test set is based on a combination of open datasets and can be used to test the capabilities of VLM and LLMs at the beginning of an interaction. 
\item By comparing different design choices on the test set, 
provide insights on whether VLM / LLMs are suitable for the problem and their current challenges.
\end{itemize}



%

\section{Related Works}
\label{related_works}

A related field to beginning an interaction 
is ``engagement," which refers to the interaction interest of a person toward a machine \cite{sidner2005explorations, bohus2014directions, anzalone2015evaluating}. The field includes variations such as non-engagement \cite{rossi2018disappearing} and human interruptibility \cite{banerjee2017temporal}.
However, most of the engagement research does not consider the robot situation (e.g., busyness or urgency of the interaction) and is primarily focused on the human side of the interaction.

Another related field is turn-taking, which tries to understand the interaction situation between the human and machine by taking into account the situation of both parties \cite{nouri2014initiative, skantze2017towards}. However, research on turn-taking begins from the state where both parties are aware of each other, thus less focused on situations where the human may be under a certain activity and unaware of the robot. 

To this end, our previous work \cite{sasabuchi2018agreeing} has worked on the beginning-of-an-interaction, which focuses on a broader view of the human-robot situation with state variety in both the human and the robot. 
However, several advancements in LLM and VLMs have occurred since then, raising the question of how these technologies fit to the problem.

The applicability of LLM and VLM has been studied in a wide range of fields, including social science \cite{wang2024survey}.
Results from \cite{gandhi2024understanding} indicate that models like GPT are able to mirror human inference patterns in Theory of Mind tests.
\cite{choi2023llms} has indicated the potential of LLMs in understanding social knowledge.
\cite{kokomind2023} shows how recent LLMs and VLMs are capable of reading verbal and non-verbal cues.
Furthermore, our study has reported the potential of GPT in understanding nuanced emotions in text \cite{wake2023bias}.
These studies on 
LLMs and VLMs indicate some promising directions for using these models to understand the situation of the human.
The question is, how do these models help generate social actions for robots, i.e., HRI.

Several recent works have focused on using pre-trained LLM/VLM models in HRI. \cite{kim2024understanding} has investigated how LLM-powered robots compare with text-based agents and voice agents. \cite{wang2024lami} has used LLM to plan human-assistive actions from multi-modal scene perception. \cite{lee2023developing} has used LLM to generate non-verbal cues in a conversation system.
\cite{verma2024theory} has studied mental-model reasoning abilities of LLMs to improve behavior synthesis in HRI.
However, research in this area is still limited \cite{kim2024survey} compared to task planning in robotics \cite{wake2024gpt}. To the best of our knowledge, no previous work has discussed the applicability of LLM/VLMs at the beginning of a human-robot interaction.



\section{Problem}
\label{problem}

The term ``interaction," especially in the context of human-machine interaction, can be viewed in different ways \cite{hornbaek2017interaction}. We will specifically focus on the term in its relation to the process of communication,
where communication is about filling the gaps between individuals and generating a consensus \cite{craig1999communication, bunt2011semantics}.
The beginning of an interaction tries to fill this first gap where one individual is trying to communicate and the other(s) may or may not have yet agreed to communicate. 

Looking from the robot's perspective in an HRI situation, there are two main situations.
One is where
the human is the one trying to communicate \cite{heenan2014designing}. In this situation, the robot must understand that the human is trying to communicate, and then choose the appropriate action (such as greet if the human is trying to communicate or ask to wait if the robot is busy).
The other situation is where
the robot is the one trying to communicate \cite{shi2013model}. In this situation, the robot must understand whether the human has agreed or is agreeing to communicate, and then choose the appropriate action (including actions where the robot should give up or postpone interacting if inappropriate).

\begin{figure}[t]
\centering
\includegraphics[width=1.0\columnwidth]{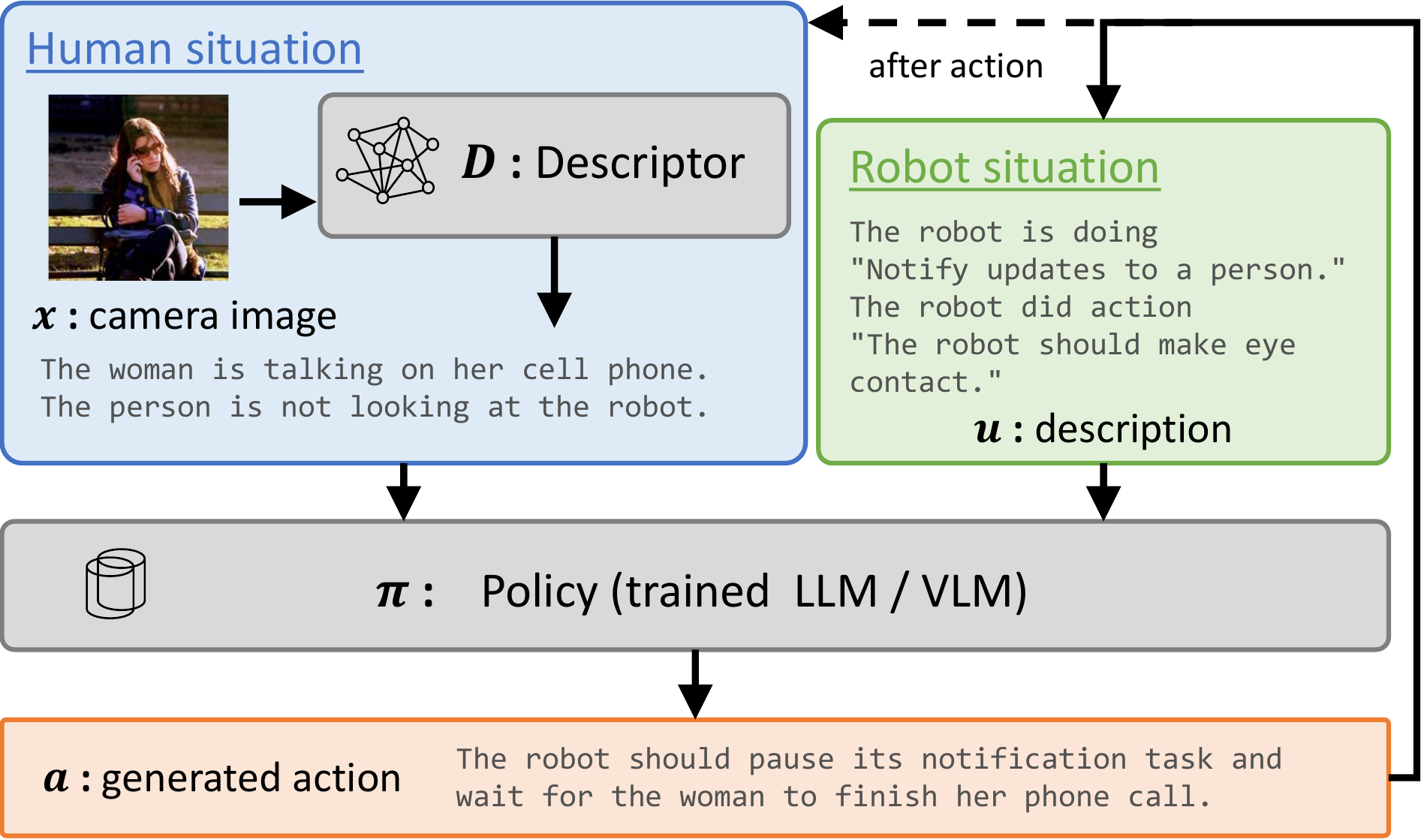}
\caption{System for beginning a human-robot interaction. Different design patterns for the descriptor and policy are tested and compared in the experiments.}
\label{fig:method}
\vspace{-3mm}
\end{figure}

Either way, the objective is for the robot to determine the action based on the situation of the human but also the situation of the robot (such as the busyness of the robot or urgency of the communication). Therefore, we can define the problem as obtaining a policy $\pi(a|s,u)$ where $a$ is a set of actions the robot can take, $s$ the situation of the human, and $u$ the situation of the robot. Since the human's situation is an indirectly extracted information obtained from the robot's sensor, we can further formalize the policy using $s=D(x)$ where $D(x)$ is a descriptor and $x$ is the sensor input (e.g., camera image), as illustrated in \figref{fig:method}.



\section{Method}
\label{method}


To decide actions at the beginning of an interaction,
the robot must obtain the policy $\pi(a|D(x),u)$. 
We will leverage LLMs and VLMs as we expect that these large models
contain some common sense about interactions.
Since recent LLMs have multi-modal capability \cite{islam2024gpt}, below we will distinguish the term LLMs and VLMs based on their input modality, where the term LLM refers to text-only input and VLM refers to a multi-modal text and image input.

There are several ways in which these models can be leveraged. One way is to replace the entire policy using a single VLM (directly obtain a policy $\pi(a|x,u)$). Another way is to replace the policy using LLMs and replace the descriptor using VLMs. A third way is to replace the policy using LLMs and implement the descriptor by combining VLMs (the descriptor $D(x)$ is a combination of descriptors like a gaze detector and a VLM image descriptor). We will compare these different system design approaches in the experiment. 



The remainder of this section goes over the detailed prompt design for leveraging the LLM and VLMs in the different design approaches.

\subsection{descriptor prompt design}

The VLM descriptor prompt is shown in \tabref{tab:description_prompt}. We design the prompt for the VLM to focus on the foreground and the activity of people in the foreground. 
The focus of the research is on beginning an interaction, thus we are interested in the human rather than the details in the background. 
We explicitly tell the VLM to \textit{``return only one sentence"} so that we obtain a focused description.

VLMs are powerful in image descriptions, however, in our early testing, we found that VLMs performed poorly to answer interaction-specific questions such as engagement signs (including gaze directions about whether the person in the image is facing toward the camera/robot). Since we were unable to obtain such domain-specific descriptions from the VLMs, we use a face detector \cite{serengil2024lightface} combined with a 6D head pose estimator \cite{6drepnet2024} to extract information about whether the person is looking at the robot or not.
That is, we programmatically generate a description ``The person is looking toward the robot" and ``The person is looking away from the robot" depending on whether the head pose is under or above a certain threshold.
In the experiments, we compare whether combining such domain-specific descriptions with general descriptions from VLMs helps to improve the LLM when generating actions. When combining, we simply concatenate the domain-specific description and the general description from the VLM.

\begin{table}[t]
\caption{Descriptor prompt.}
\vspace{-3mm}
\label{tab:description_prompt}
\begin{center}
\begin{tabular}{|l|p{0.8\linewidth}|}
\hline
user & Please describe what the people are doing. Do not describe the background. Do not describe the room. Do not describe what the people are wearing and focus only on what they are doing. Return only one sentence.  \\
\hline
\end{tabular}
\end{center}
\end{table}

\subsection{policy prompt design}

The prompt for the LLM policy is shown in \tabref{tab:llm_policy_prompt}. 
Although we do not bound the returned actions in the form of action sets, we will bound the form factor of the robot to produce consistent and comparable results. That is, we explicitly mention that the robot is a humanoid and is capable of verbal actions. Without this notation, inconsistent actions such as \textit{``produce a beep sound"} or actions that are unrelated to the interaction such as \textit{``scan with a lidar"} can be generated due to attention to the term \textit{``robot"}.

\begin{table}[t]
\caption{LLM policy prompt.}
\vspace{-5mm}
\label{tab:llm_policy_prompt}
\begin{center}
\begin{tabular}{|l|p{0.8\linewidth}|}
\hline
user & I will provide a sentence about a human-humanoid interaction situation. Your job is to figure out how the humanoid robot should behave in this situation by returning a dictionary with the keys ``action" and ``reason". Please keep in mind that some people may feel uncomfortable interacting with the robot. If the robot requires some verbal action, also come up with an appropriate phrase.

Below is an example:

- My input: ``The robot is waiting to help a person. A person is approaching the robot. The person is looking at the robot."

- Your output: \{``action": ``The robot should turn towards the person and make eye contact.", ``reason": ``The robot should acknowledge the person's presence. The robot should not speak yet as the person may feel uncomfortable and not engaged yet."\}

Understood? \\
\hline
assistant & Understood! Please provide the sentence about the humanoid-robot interaction situation, and I'll determine the appropriate behavior for the humanoid robot. \\
\hline
user & The robot is doing \textit{U}. \textit{X}. \\
\hline
assistant & \{``action": \textit{A}, ``reason": \textit{Y} \} \\
\hline
user & The robot is doing \textit{U}. The robot did action: \textit{A}. After this action: \textit{X}. \\
\hline
\end{tabular}
\end{center}
\end{table}

In addition, a notion that \textit{``some people may feel uncomfortable interacting with the robot"} is provided in the prompt to explicitly guide that the situation is about the beginning of an interaction and that there is uncertainty on whether the human is agreeing to interact with the robot. Furthermore, general tricks such as providing an example and asking to return the reasoning are included in the prompt to improve performance.

After the above prompt, the situation (a concatenated text of the human and robot situation) is passed to the LLM for the LLM to generate an action. 
It is important to note that the LLM policy 
maybe triggered multiple times until the human and the robot 
agree to communicate, 
such as observing the situation more before begin speaking.

In order to achieve the above, an updated situation is passed to the LLM. 
Every time the LLM policy returns an action, the next text passed to the LLM policy will start with \textit{``The robot is doing U. The robot did action: A. After this action: X"} where \textit{U} is the robot's situation, \textit{A} is the action returned by the LLM policy in the previous round, and \textit{X} is the new situation obtained after the action was performed.
The robot's situation is passed every time to decrease the chances of the LLM losing attention toward the situation.

The prompt for the VLM policy (as an alternative to the LLM policy) is mostly the same except: (1) we only provide the robot situation \textit{U} as text and instead provide the image directly for the human situation, (2) we ask the policy to also return the image description.
The prompt for the VLM policy is shown in \tabref{tab:vlm_policy_prompt} with the difference from the LLM policy highlighted in bold.

\begin{table}[t]
\caption{VLM policy prompt.}
\vspace{-5mm}
\label{tab:vlm_policy_prompt}
\begin{center}
\begin{tabular}{|l|p{0.8\linewidth}|}
\hline
user & I will provide a sentence about a human-humanoid interaction situation \textbf{and an image of the current human's situation obtained from the robot's camera.} Your job is to figure out how the humanoid robot should behave in this situation by returning a dictionary with the keys ``action," ``reason," and \textbf{``image\_description"}. Please keep in mind that some people may feel uncomfortable interacting with the robot. If the robot requires some verbal action, also come up with an appropriate phrase.

Below is an example:

- My input: ``The robot is waiting to help a person."

- Your output: \{``action": ``The robot should turn towards the person and make eye contact.", ``reason": ``The robot should acknowledge the person's presence. The robot should not speak yet as the person may feel uncomfortable and not engaged yet.", \textbf{``image\_description": ``A person is walking in an office looking a the robot showing a chance of engagement."}\}

Understood? \\
\hline
assistant & Understood! Please provide the sentence about the humanoid-robot interaction situation, and I'll determine the appropriate behavior for the humanoid robot. \\
\hline
user & The robot is doing \textit{U}. \\
\hline
assistant & \{``action": \textit{A}, ``reason": \textit{Y}, ``image\_description": \textit{Z} \} \\
\hline
user & The robot is doing \textit{U}. The robot did action: \textit{A}. \textbf{The image was taken after this action.} \\
\hline
\end{tabular}
\end{center}
\end{table}


\section{Test set}
\label{testset}

\begin{figure}[t]
\centering
\includegraphics[width=\columnwidth]{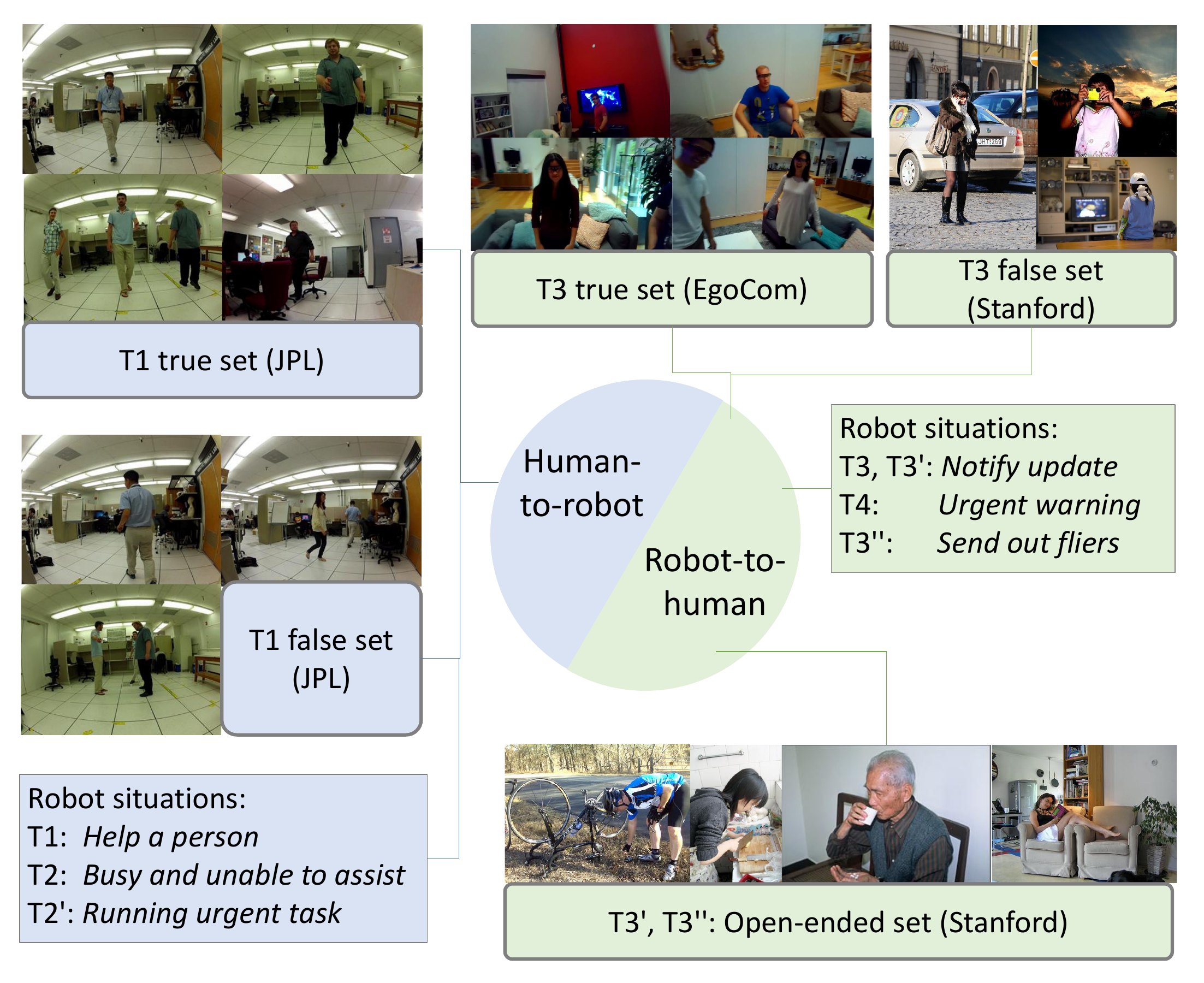}
\caption{Overview of the test set with example videos/images.}
\label{fig:testset}
\vspace{-3mm}
\end{figure}

\begin{table}[t]
\caption{List of video/image files used from each dataset. For the Stanford dataset the first image in the category is used unless a number is provided in paranthesis. Images containing multiple people were avoided for the Standford dataset.}
\vspace{-3mm}
\label{tab:testset}
\begin{center}
\begin{tabular}{|l|p{0.6\linewidth}|}
\hline
JPL (true set) & c1, c19, c24, c30 (second-half), c35, c42, c50 \\
\hline
JPL (false set) & c1, c3, c30 (first-half) \\
\hline
EgoCom (true set) & 090, 093, 101, 116 \\
\hline
Stanford (false set) & phoning(002), taking photos(004), watching TV \\
\hline
Stanford (open-ended) & applauding(003), blowing bubbles, brushing teeth, cleaning the floor(002), cooking, cutting trees, cutting vegetables, drinking, feeding a horse, fishing, fixing a bike(002), gardening, holding an umbrella(002), looking through a microscope(002), playing guitar, pouring liquid(002), pushing a cart(005), reading, smoking, texting message(003), using a computer, walking the dog, washing dishes, waving hands(005), writing on a board(003) \\
\hline
\end{tabular}
\end{center}
\end{table}

As explained in \secref{problem},
there are two main situations depending on whether the human or the robot is trying to communicate. In addition, 
there are two main problems to solve, understanding the human situation and deciding the robot action. 
To understand whether LLMs and VLMs are applicable to these situations and problems, we raise the following questions:
\begin{itemize}
\item (RQ1) When the human is trying to communicate, will these models be able to understand that the human is trying to communicate?
\item (RQ2) When the human is trying to communicate, will these models output different actions depending on the robot's situation (busyness of the task)?
\item (RQ3) When the robot is trying to communicate, will these models be able to understand whether the human is in a state that is okay to speak to?
\item (RQ4) When the robot is trying to communicate, will these models output different actions depending on the robot's situation (urgency of the task)?
\end{itemize}

To answer these questions,
we need a test set containing videos/images of a human taken from a robot's camera with the following situations: 
\begin{itemize}
\item (T1) A test set including a true set where the human is trying to communicate and a false set where the human is not trying to communicate. 
In this test, the model should return agreeing actions (such as speak) only on the true set.
\item (T2) Using the above true set, change the busyness of the robot's situation. In this test, the model should return differing actions based on the situational difference.
\item (T3) A test set including a true set where the human is in an okay to speak state and a false set where the human is in a not okay to speak state. In this test, 
the robot should return agreeing actions (such as speak) only on the true set.
\item (T4) Using the above false set, change the urgency of the robot's situation. In this test, the model should return differing actions based on the situational difference.
\end{itemize}

Unfortunately, there are very few datasets/test sets in HRI that are publicly available with video/image data, and none specifically match our purpose. 
Therefore, we mix multiple datasets. 

For T1, we leverage the JPL interaction dataset \cite{ryoo2013first} which is one of the few publicly available video dataset in HRI. 
For our testing purposes, we organized the videos into three classes: videos where the human(s) engaged with the robot, the human(s) did not engage with the robot, and videos containing non-general human interaction behaviors such as punching the robot. We use the first class as the true set, the second class as the false set, and exclude the third class. 
As our focus is on situational variety,
we remove duplicate videos that only differ in the subject or 
contain the same activity pattern. 
This leaves us with seven videos for the true set and three videos for the false set as listed in \tabref{tab:testset}. We further split the videos into images with a 1.5 seconds interval where each image is fed one at a time as input $x$ to generate actions. The input $u$ is fixed to \textit{``Help a person if the person needs assistance."}

For T2, we simply change the text description $u$ to \textit{``Busy with a task and unable to assist."} The robot's inability to assist is explicitly mentioned, and it is expected that the robot will avoid or postpone the interaction.

For T3, many HRI research in the area focuses on motion generation and lacks datasets to test action decisions. Therefore, as an alternative, we leverage videos from human-human interaction datasets taken from an ego-centric view. We use the publicly available EgoCom dataset \cite{egocom2020} which contain such videos applicable to the true set situation. From the dataset, we choose videos where the participant of the first-person view speaks to the other person/people but exclude videos where there is too much first-person head movement that goes back and forth between two people, and videos where the faces are not properly captured due to the eye-level of the camera. This leaves us with four valid videos listed in \tabref{tab:testset} but with a variety of activity patterns. We further split the videos into images 
and choose two non-blurred frames with more than 0.75 seconds interval in between. The input $u$ is fixed to \textit{``Notify updates to a person."} 

Since the EgoCom dataset does not include situations where the human is in a state not okay to speak, we further leverage images from the Standford 40 Activity Dataset \cite{yao2011human} which contain a variety of human activities such as phone calling. Of the 40 activity categories, we use images from the category phoning, taking photos, and watching TV for the false set as shown in \figref{fig:testset}. Although the dataset includes more activity categories (e.g., drinking), the appropriateness of disrupting these activities was arguable. For these activities, we create a third open-ended set which is explained later in the section.

For T4, we simply change the text description $u$ to \textit{``Report an urgent warning that must be reported to the person,"} and the human situation is an image of either phoning, taking photos, or watching TV. The report is explicitly stated as a must, and it is expected that the robot will try to speak regardless of the human situation.

The above tests have a clear answer to the expected action output. However, there are situations where there is no clear answer to the correct action. Some human activities could be disrupted, but waiting for the activity to finish may also be acceptable. This raises the additional question of how the LLM/VLM models would respond to situations where the appropriate action is open-ended. Will the models prioritize the human situation or the robot situation? To address this question, we create the following additional test sets:
\begin{itemize}
\item (T2') An open-ended robot situation test where the robot's availability of the interaction is not stated.
\item (T3') An open-ended human situation test where the okay-to-speak state of the human is unlabeled.
\item (T3'') An open-ended human and robot situation test where the okay-to-speak state of the human is unlabeled and the necessity of the robot's interaction is not stated.
\end{itemize}

For T2', we test the case where $u$ is \textit{``Running an urgently requested task by another person."} This description only mentions that a busy task is going on but leaves room for the model to decide whether to assist the human who is trying to communicate.

For T3', we leverage images from the remaining activity categories in the Standford 40 Activity Dataset. However, since the purpose is to understand responses of these models against situational variance, we removed activities that are similar to another such as ``playing violin" which is similar to ``playing guitar." 
We also excluded all sports activities to focus on situations where the person is assumed to remain mostly static while the robot tries to communicate. This leaves us with 25 activities listed in \tabref{tab:testset}. For input $u$, we use the same text description as T3.

For T3'', we use the images from T3' but change $u$ to \textit{``Send out fliers to only the people who are willing to accept."} This provides a vague description about whether the robot should or should not interact with the human, but instead provides a concrete description about the robot's task.


\section{Experiments}
\label{experiment}

\begin{figure*}[t]
\centering
\includegraphics[width=1.8\columnwidth]{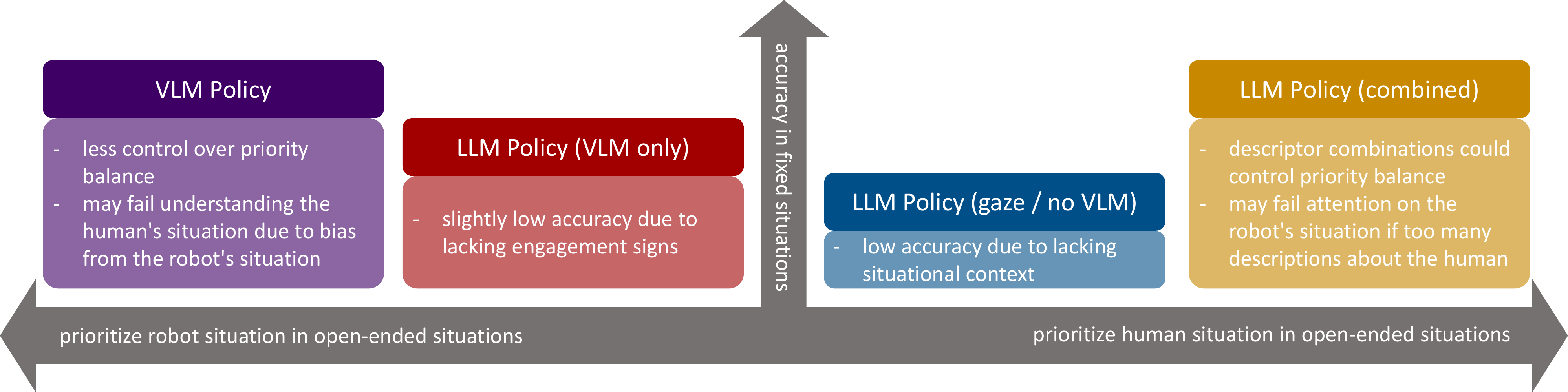}
\caption{Summary of findings from results. Parentheses indicate descriptor type. The VLM policy and the LLM policy with the combined descriptor generate the most appropriate actions in non-open-ended situations. However, for open-ended situations, the VLM policy prioritizes the robot situation and the LLM policy with the combined descriptor prioritizes the human situation and have challenges in balancing situational priority.}
\label{fig:results}
\vspace{-3mm}
\end{figure*}

Using the test set in \secref{testset}, we compare the following setups:
(A) policy is LLM but the descriptor is gaze observations (no VLM),
(B) policy is LLM and the descriptor is VLM, (C) policy is LLM and the descriptor is a combination of VLM and gaze observations, (D) entire policy is a single VLM.
Approach A was not explained in \secref{method}, but this approach acts as a baseline comparison to observe the effectiveness of VLMs compared to less descriptive descriptors. 

For the LLM/VLM policy we use Azure OpenAI GPT-4o (2024-08-06 model) and for the VLM descriptor we use Phi-3 Vision. Both models are publicly accessible and are one of the models with the highest performance \cite{islam2024gpt, abdin2024phi}. We chose the Phi-3 Vision model for the VLM descriptor as it is capable to run on an edge machine and decreases latency.

Realtime HRI requires decisions in one to two seconds. 
Due to these model choices, the setups A, B, C can be practically applied on a real robot system with one to 1.5 seconds response time; however, the setup for D requires three to five seconds response time 
and 
results are subject to \textit{if} the approach was applicable on a real robot system.

\subsection{T1 results}

\begin{table}[b]
\caption{T1, T2, T2' results. Number of videos where the model returned an agreeing action (assisting the human) out of the total number of test videos. Higher is better for the true set. Lower is better for the false set and T2. T2' numbers are to compare with the other test conditions.}
\vspace{-3mm}
\label{tab:T1}
\begin{center}
\begin{tabular}{|l|c|c|c|c|}
\hline
 & A & B & C & D  \\
\hline
T1 true set & 7/7 & 4/7 & 7/7 & 7/7 \\
\hline
T1 false set & 2/3 & 2/3 & 0/3 & 2/3 \\
\hline
T2 & 1/7 & 1/7 & 4/7 & 0/7 \\
\hline
T2' & 7/7 & 2/7 & 7/7 & 2/7 \\
\hline
\end{tabular}
\end{center}
\end{table}

The results for T1 are shown in \tabref{tab:T1}, and examples of the actions generated are shown in \tabref{tab:examples}. For the true set, most approaches were able to understand the situation of the human, but to our surprise, the VLM descriptor performed the worst in this set. The reason for this was clear when observing the responses 
as shown in \tabref{tab:reasons}. The LLM policy had misinterpreted the description ``walking" from the VLM descriptor as a person passing by rather than a person walking towards the robot. All other approaches did not have such vagueness as the image was passed directly to the policy or as was combined with gaze observations. 

For the false set, where lower numbers are better, approach C with the combined descriptor was the only model that was able to understand that the person was not requiring assistance. Again, we are able to observe the reasons behind from the responses of each approach as shown in \tabref{tab:reasons}. While approaches A and B failed due to the misinterpretation of the generated descriptions, approach D had an interesting result where, 
the VLM had a high prior about the image based on the input text. That is, since the text included information about the robot's role in assisting people, the VLM tried to interpret the image as a picture of someone who is needing assistance regardless of their actual situation.

\subsection{T2, T2' results}

The results for T2 are shown in \tabref{tab:T1}. When the robot's situation was explicitly mentioned that it was unable to assist, most approaches returned actions different from T1 as shown in \tabref{tab:examples} and avoided immediate assistance. However, in contrast to the first experiment, approach C with the combined descriptor performed worst. As shown in \tabref{tab:reasons} most of the failure cases were due to not paying attention to the robot situation. Approach C performed worst possibly as the attention was the most difficult to maintain due to having multiple descriptions of the human situation.

For T2', when the availability for assistance was not explicitly mentioned, we observe some interesting results. For all approaches, the policies decided to help more compared to T2. 
However, compared to the T1 situation, 
the VLM policy provided significantly less interaction, 
From the results, we can observe that when information about engagement signs (gaze observations) are explicitly provided, the policies tend to prioritize engagement with the human, but when the signs are not provided, the policies tend to prioritize the task described in the robot's situation.

\subsection{T3, T4 results}

\begin{table}[t]
\caption{T3, T4 results. Number of videos/images where the model returned an agreeing action (speak) instead of waiting out of the total number of test videos/images. Higher is better for the true set and T4. Lower is better for the false set.}
\vspace{-3mm}
\label{tab:T3}
\begin{center}
\begin{tabular}{|l|c|c|c|c|}
\hline
 & A & B & C & D  \\
\hline
T3 true set & 4/4 & 4/4 & 4/4 & 4/4 \\
\hline
T3 false set & 3/3 & 0/3 & 0/3 & 0/3 \\
\hline
T4 & 3/3 & 2/3 & 3/3 & 3/3 \\
\hline
\end{tabular}
\end{center}
\end{table}

The results for T3 are shown in \tabref{tab:T3} and the generated actions in \tabref{tab:examples}. As expected, the descriptor with the gaze-observation-only approach lacks situational context and fails by trying to catch attention regardless of whether the human is calling over a phone, whereas, all other approaches tried to wait until the human finishes their activity. 

The results for T4 are also shown in \tabref{tab:T3}. The results indicate that, depending on the situation, these models are able to prioritize the robot's situation over the human's situation when there is a need of urgency (the models will try to catch the human's attention even if they are talking over the phone if the report is urgent). However, there was one case with the VLM descriptor where the policy did not try to report to the human due to missing attention toward the robot's situation and focusing only on the human's activity.

\subsection{T3', T3'' results}

\begin{table}[t]
\caption{T3', T3'' results. Number of images where the model returned an agreeing action (speak) out of the total number of test videos.}
\vspace{-4mm}
\label{tab:T5}
\begin{center}
\begin{tabular}{|l|c|c|c|c|}
\hline
 & A & B & C & D  \\
\hline
T3' & 25/25 & 5/25 & 10/25 & 12/25 \\
\hline
T3'' & 15/25 & 10/25 & 7/25 & 18/25 \\
\hline
\end{tabular}
\end{center}
\end{table}

The results for T3' and T3'' are shown in \tabref{tab:T5}. The results of T3' indicates that when the situation is open-ended, the generated actions become a mix of waiting and catching attention, except for approach A which always tried to catch attention for the same reason as T3. The results of T3' are not a surprise, but we observe interesting results when compared these results with T3''.

The difference between T3' and T3'' is that T3'' further adds open-endedness to the robot situation. When this is added, approaches A and C decrease in the number of times the policy generated speaking actions, while approaches B and D had an increase. Observing the reasons for the action decision as shown in \tabref{tab:reasons}, it is apparent that the latter two approaches prioritized the robot's situation (task of sending out the flier) over the human's activity. This is somewhat similar to the observations in T2'. When information about engagement signs is not explicitly provided, the policies tend to prioritize the task described in the robot's situation.

\section{Conclusions}
\label{conclusions}

A summary of our findings is illustrated in \figref{fig:results}.
Although results are limited to the tested GPT-4O and Phi-3 Vision model, the results provide us with a lower bound of what LLM/VLMs are capable of, as well as
insights on whether these models are suitable for the problem, and how different inputs may affect the generation of the actions.
From the experiments, it is clear that the LLM and VLM policies are capable of understanding the human situation and generating actions by incorporating both the human and robot situation. It is also clear that VLMs helps better understand the human situation, especially when the robot is trying to initiate the interaction. As shown in \tabref{tab:examples}, 
the LLM-based policy has the advantage of returning detailed actions such as generating speech content but also generating detailed wait conditions such as \textit{``until the woman finishes her phone call."}

Despite these advantages, we also observed some limitations. There remains a challenge to balance between the human situation and the robot situation when the possible actions are open-ended. One situation can be prioritized against the other from slight differences in how the human situation is explained, or depending on the robot's task. Results from the open-ended tests indicate that the more information is provided about the engagement signs (gaze observations), the more attention is paid to the human situation. A single VLM policy, while having fewer loss in information, may not necessarily weigh importance on engagement signs, and thus an LLM policy combined with different descriptors may add more flexibility in manipulating this balance. Nevertheless, the results indicate that when the situation is less open-ended and when the desired actions are clear, recent LLMs and VLMs are able to provide appropriate actions at the beginning of a human-robot interaction. 

\begin{table*}[t]
\caption{Examples of generated robot actions for the tests (shortened from actual response).}
\vspace{-4mm}
\label{tab:examples}
\begin{center}
\begin{tabular}{|l|p{0.2\linewidth}|p{0.2\linewidth}|p{0.2\linewidth}|p{0.2\linewidth}|}
\hline
 & A) LLM policy with gaze & B) LLM policy with VLM & C) LLM policy with combined & D) VLM policy  \\
\hline
T1
& observe the person without being intrusive; acknowledge the person's attention by nodding slightly; maintain eye contact and wait patiently; say 'Hello! Do you need any assistance?'
& observe the man while remaining stationary; approach the man slowly and stop at a respectful distance; politely ask if assistance is needed
& remain stationary and monitor the situation discreetly; turn slightly towards the woman who is looking at the robot; make eye contact and offer assistance verbally
& observe the person for any signs of needing assistance, such as looking around or appearing confused; approach the person and ask if they need assistance \\
\hline
T2
& continue with its task without initiating interaction; say 'Hello, I see you are looking at me. I am currently busy, but I will be available to assist you shortly.'
& maintain its task but keep an eye on the man; say 'Hello, I am currently busy with a task. If you need assistance, I will be available shortly.'
& continue with its task and avoid interrupting the man; say 'Hello, is there anything I can assist you with once I finish my current task?'; say 'Hello, how can I assist you today?'
& display a polite message indicating it is currently busy; display a message indicating it will be available soon \\
\hline
T2'
& briefly pause the task; acknowledge the person's attention with a brief nod; say 'Do you need any help, or should I continue with the current task?'
& continue focusing on the urgent task without further interaction; say 'Hello, I'll be with you shortly.'
& maintain eye contact briefly and offer a polite nod; say 'It looks like you might need some help. Please let me know how I can assist you.'
& focus on completing the task efficiently and avoid unnecessary interactions; pause briefly to assess if the person needs assistance or is signaling something important \\
\hline
T3 / T3'
& alert the person with a soft auditory signal; say 'Hello! I have some updates for you. Would you like to hear them now?'
& wait until the woman finishes her phone call; say 'Hello, I have some updates for you. Would you like to hear them now?'
& wait for the woman to finish her phone call; say 'Hello, I have some updates for you. Would you like to hear them now?'
& approach the person quietly and wait for a pause in their activity; approach the person and use a polite tone to deliver the updates \\
\hline
T4
& say 'Excuse me, I have an urgent warning to report.'
& say 'Excuse me, I have an urgent warning that needs your attention.'
& say 'Excuse me, I have an urgent warning that needs your attention.'
& approach the person calmly and make a sound to get their attention \\
\hline
T3''
& wait patiently and avoid initiating interaction; say 'Hello! Would you like to receive a flier?'
& remain stationary and silent, waiting for the woman to acknowledge its presence; approach the man slowly and stop at a respectful distance; offer a flier with a friendly gesture
& wait quietly and observe the woman until she finishes her task or acknowledges the robot; 
raise its hand with a flier and ask if the woman would like to receive one
& refrain from offering a flier and continue observing for interested individuals; approach the person slowly and non-intrusively, ensuring it is within their line of sight \\ 
\hline
\end{tabular}
\end{center}
\end{table*}

\begin{table*}[t]
\caption{Examples of returned reasons for the robot action decisions including failure cases.}
\vspace{-3mm}
\label{tab:reasons}
\begin{center}
\begin{tabular}{|l|p{0.2\linewidth}|p{0.2\linewidth}|p{0.2\linewidth}|p{0.2\linewidth}|}
\hline
 & A) LLM policy with gaze & B) LLM policy with VLM & C) LLM policy with combined & D) VLM policy  \\
\hline
T1 true
& The person is now looking at the robot, which may indicate they are seeking help or interaction.
& (FAIL) Since the man is walking, he may be moving towards a specific destination or task.
& Since the person is looking at the robot, they might be considering whether they need assistance.
& The person's gesture might indicate they are about to communicate or request something. \\
\hline
T1 false
& (FAIL) Since the person is still not looking at the robot, they might not be aware of its presence or the assistance it can offer.
& (FAIL) The change in body language from gesturing to standing might indicate a pause or conclusion in their conversation, providing an opportunity for the robot to offer help.
& The woman is still not indicating any need for assistance, as she is not looking at the robot or showing signs of requiring help.
& (FAIL) A person is walking through an office environment, possibly looking for something or someone to assist them. \\
\hline
T2
& (FAIL) The person is now looking at the robot, indicating potential interest or need for assistance.
& (FAIL) The man's movement in the office might indicate he is looking for something or someone to assist him.
& (FAIL) The man is now looking at the robot, indicating he may need assistance or is interested in interacting.
& Communicating its unavailability helps manage expectations and prevents frustration. \\
\hline
T2'
& Acknowledging the person with a simple greeting allows the robot to be polite and responsive, while also providing an opportunity for the person to express any immediate needs or concerns.
& Acknowledging the man's presence is polite and shows awareness of the environment, but the robot should not engage further unless approached, to maintain focus on the urgent task and avoid unnecessary interruptions.
& Since the person is looking at the robot, a brief verbal acknowledgment can address their attention without significantly interrupting the task. This approach balances politeness with the need to prioritize the urgent task.
& The task is urgent, so the robot should prioritize completing it over engaging with others in the environment. \\
\hline
T3'
& Since the person is not looking at the robot, a gentle auditory signal can capture their attention without being intrusive.
& Waiting for a natural pause or acknowledgment ensures the man is ready to receive the information.
& The robot should verbally notify the person, but it should do so politely and without urgency, as the person is currently engaged in an activity.
& The person appears to be engaged in a leisure activity, so the robot should be considerate and not disruptive. \\
\hline
T3''
& Since the person is not looking at the robot, they may not be ready or willing to engage.
& By asking for permission, the robot respects the man's choice and comfort level with receiving a flier.
& Approaching someone who is not looking at the robot might startle or discomfort them.
& The person appears to be focused and may not notice the robot. Asking politely ensures the person is willing to accept the flier. \\
\hline
\end{tabular}
\end{center}
\end{table*}

\ifconfletter
\bibliographystyle{IEEEtran}
\else
\bibliographystyle{unsrt}
\fi
\bibliography{root}

\begin{thebibliography}{10}

\bibitem{sasabuchi2018agreeing}
Kazuhiro Sasabuchi, Katsushi Ikeuchi, and Masayuki Inaba.
\newblock Agreeing to interact: Understanding interaction as human-robot goal conflicts.
\newblock In {\em Companion of the 2018 ACM/IEEE International Conference on Human-Robot Interaction}, pages 21--28, 2018.

\bibitem{binz2023using}
Marcel Binz and Eric Schulz.
\newblock Using cognitive psychology to understand gpt-3.
\newblock {\em Proceedings of the National Academy of Sciences}, 120(6):e2218523120, 2023.

\bibitem{sidner2005explorations}
Candace~L Sidner, Christopher Lee, Cory~D Kidd, Neal Lesh, and Charles Rich.
\newblock Explorations in engagement for humans and robots.
\newblock {\em Artificial Intelligence}, 166(1-2):140--164, 2005.

\bibitem{bohus2014directions}
Dan Bohus, Chit~W Saw, and Eric Horvitz.
\newblock Directions robot: in-the-wild experiences and lessons learned.
\newblock In {\em Proceedings of the 2014 international conference on Autonomous agents and multi-agent systems}, pages 637--644, 2014.

\bibitem{anzalone2015evaluating}
Salvatore~M Anzalone, Sofiane Boucenna, Serena Ivaldi, and Mohamed Chetouani.
\newblock Evaluating the engagement with social robots.
\newblock {\em International Journal of Social Robotics}, 7:465--478, 2015.

\bibitem{rossi2018disappearing}
Silvia Rossi, Giovanni Ercolano, Luca Raggioli, Emanuele Savino, and Martina Ruocco.
\newblock The disappearing robot: an analysis of disengagement and distraction during non-interactive tasks.
\newblock In {\em 2018 27th IEEE international symposium on robot and human interactive communication (RO-MAN)}, pages 522--527. IEEE, 2018.

\bibitem{banerjee2017temporal}
Siddhartha Banerjee and Sonia Chernova.
\newblock Temporal models for robot classification of human interruptibility.
\newblock In {\em Proceedings of the 16th Conference on Autonomous Agents and MultiAgent Systems}, pages 1350--1359, 2017.

\bibitem{nouri2014initiative}
Elnaz Nouri and David Traum.
\newblock Initiative taking in negotiation.
\newblock In {\em Proceedings of the 15th Annual Meeting of the Special Interest Group on Discourse and Dialogue (SIGDIAL)}, pages 186--193, 2014.

\bibitem{skantze2017towards}
Gabriel Skantze.
\newblock Towards a general, continuous model of turn-taking in spoken dialogue using lstm recurrent neural networks.
\newblock In {\em Proceedings of the 18th Annual SIGdial Meeting on Discourse and Dialogue}, pages 220--230, 2017.

\bibitem{wang2024survey}
Lei Wang, Chen Ma, Xueyang Feng, Zeyu Zhang, Hao Yang, Jingsen Zhang, Zhiyuan Chen, Jiakai Tang, Xu~Chen, Yankai Lin, et~al.
\newblock A survey on large language model based autonomous agents.
\newblock {\em Frontiers of Computer Science}, 18(6):186345, 2024.

\bibitem{gandhi2024understanding}
Kanishk Gandhi, Jan-Philipp Fr{\"a}nken, Tobias Gerstenberg, and Noah Goodman.
\newblock Understanding social reasoning in language models with language models.
\newblock {\em Advances in Neural Information Processing Systems}, 36, 2024.

\bibitem{choi2023llms}
Minje Choi, Jiaxin Pei, Sagar Kumar, Chang Shu, and David Jurgens.
\newblock Do llms understand social knowledge? evaluating the sociability of large language models with socket benchmark.
\newblock {\em arXiv preprint arXiv:2305.14938}, 2023.

\bibitem{kokomind2023}
Weiyan Shi, Liang Qiu, Dehong Xu, Pengwei Sui, Pan Lu, and Zhou Yu.
\newblock Kokomind: Can large language models understand social interactions?, July 2023.

\bibitem{wake2023bias}
Naoki Wake, Atsushi Kanehira, Kazuhiro Sasabuchi, Jun Takamatsu, and Katsushi Ikeuchi.
\newblock Bias in emotion recognition with chatgpt.
\newblock {\em arXiv preprint arXiv:2310.11753}, 2023.

\bibitem{kim2024understanding}
Callie~Y Kim, Christine~P Lee, and Bilge Mutlu.
\newblock Understanding large-language model (llm)-powered human-robot interaction.
\newblock In {\em Proceedings of the 2024 ACM/IEEE International Conference on Human-Robot Interaction}, pages 371--380, 2024.

\bibitem{wang2024lami}
Chao Wang, Stephan Hasler, Daniel Tanneberg, Felix Ocker, Frank Joublin, Antonello Ceravola, Joerg Deigmoeller, and Michael Gienger.
\newblock Lami: Large language models for multi-modal human-robot interaction.
\newblock In {\em Extended Abstracts of the CHI Conference on Human Factors in Computing Systems}, pages 1--10, 2024.

\bibitem{lee2023developing}
Yoon~Kyung Lee, Yoonwon Jung, Gyuyi Kang, and Sowon Hahn.
\newblock Developing social robots with empathetic non-verbal cues using large language models.
\newblock {\em arXiv preprint arXiv:2308.16529}, 2023.

\bibitem{verma2024theory}
Mudit Verma, Siddhant Bhambri, and Subbarao Kambhampati.
\newblock Theory of mind abilities of large language models in human-robot interaction: An illusion?
\newblock In {\em Companion of the 2024 ACM/IEEE International Conference on Human-Robot Interaction}, pages 36--45, 2024.

\bibitem{kim2024survey}
Yeseung Kim, Dohyun Kim, Jieun Choi, Jisang Park, Nayoung Oh, and Daehyung Park.
\newblock A survey on integration of large language models with intelligent robots.
\newblock {\em Intelligent Service Robotics}, 17(5):1091--1107, 2024.

\bibitem{wake2024gpt}
Naoki Wake, Atsushi Kanehira, Kazuhiro Sasabuchi, Jun Takamatsu, and Katsushi Ikeuchi.
\newblock Gpt-4v (ision) for robotics: Multimodal task planning from human demonstration.
\newblock {\em IEEE Robotics and Automation Letters}, 2024.

\bibitem{hornbaek2017interaction}
Kasper Hornb{\ae}k and Antti Oulasvirta.
\newblock What is interaction?
\newblock In {\em Proceedings of the 2017 CHI conference on human factors in computing systems}, pages 5040--5052, 2017.

\bibitem{craig1999communication}
Robert~T Craig.
\newblock Communication theory as a field.
\newblock {\em Communication theory}, 9(2):119--161, 1999.

\bibitem{bunt2011semantics}
Harry Bunt.
\newblock The semantics of dialogue acts.
\newblock In {\em Proceedings of the Ninth International Conference on Computational Semantics (IWCS 2011)}, 2011.

\bibitem{heenan2014designing}
Brandon Heenan, Saul Greenberg, Setareh Aghel-Manesh, and Ehud Sharlin.
\newblock Designing social greetings in human robot interaction.
\newblock In {\em Proceedings of the 2014 conference on Designing interactive systems}, pages 855--864, 2014.

\bibitem{shi2013model}
Chao Shi, Masahiro Shiomi, Christian Smith, Takayuki Kanda, and Hiroshi Ishiguro.
\newblock A model of distributional handing interaction for a mobile robot.
\newblock In {\em Robotics: science and systems}, volume~10. Berlin, Germany, 2013.

\bibitem{islam2024gpt}
Raisa Islam and Owana~Marzia Moushi.
\newblock Gpt-4o: The cutting-edge advancement in multimodal llm.
\newblock {\em Authorea Preprints}, 2024.

\bibitem{serengil2024lightface}
Sefik Serengil and Alper Ozpinar.
\newblock A benchmark of facial recognition pipelines and co-usability performances of modules.
\newblock {\em Journal of Information Technologies}, 17(2):95--107, 2024.

\bibitem{6drepnet2024}
Thorsten Hempel, Ahmed~A. Abdelrahman, and Ayoub Al-Hamadi.
\newblock Toward robust and unconstrained full range of rotation head pose estimation.
\newblock {\em IEEE Transactions on Image Processing}, 33:2377--2387, 2024.

\bibitem{ryoo2013first}
M.~S. Ryoo and L.~Matthies.
\newblock First-person activity recognition: What are they doing to me?
\newblock In {\em IEEE Conference on Computer Vision and Pattern Recognition (CVPR)}, Portland, OR, June 2013.

\bibitem{egocom2020}
Curtis~G. {Northcutt}, Shengxin {Zha}, Steven {Lovegrove}, and Richard {Newcombe}.
\newblock Egocom: A multi-person multi-modal egocentric communications dataset.
\newblock {\em IEEE Transactions on Pattern Analysis and Machine Intelligence}, pages 1--12, 2020.

\bibitem{yao2011human}
Bangpeng Yao, Xiaoye Jiang, Aditya Khosla, Andy~Lai Lin, Leonidas Guibas, and Li~Fei-Fei.
\newblock Human action recognition by learning bases of action attributes and parts.
\newblock In {\em 2011 International conference on computer vision}, pages 1331--1338. IEEE, 2011.

\bibitem{abdin2024phi}
Marah Abdin, Jyoti Aneja, Hany Awadalla, Ahmed Awadallah, Ammar~Ahmad Awan, Nguyen Bach, Amit Bahree, Arash Bakhtiari, Jianmin Bao, Harkirat Behl, et~al.
\newblock Phi-3 technical report: A highly capable language model locally on your phone.
\newblock {\em arXiv preprint arXiv:2404.14219}, 2024.

\end{thebibliography}

%








\end{document}